\documentclass[aps,prl,superscriptaddress,reprint,showpacs,longbibliography,amsmath,amssymb]{revtex4-1}
\usepackage[utf8]{inputenc}
\usepackage[T1]{fontenc}
\usepackage{microtype}
\usepackage{graphicx}
\usepackage{mathbbol}
\usepackage{amssymb}
\usepackage{comment}
\usepackage{gensymb}
\usepackage[usenames,dvipsnames,svgnames,table]{xcolor}
\usepackage[colorlinks,linkcolor=blue,citecolor=blue,urlcolor=blue]{hyperref}

\def\equationautorefname~#1\null{Eq.~(#1)\null}

\begin{document}
	
	\title{Optomechanically Induced Birefringence and Faraday Effect}
	
	\author{Robert Duggan}
	\thanks{R. D. and J. d. P. contributed equally to this work.}
	\affiliation{Department of Electrical and Computer Engineering, The University of Texas at Austin, Austin, TX 78712, USA}
	\author{Javier del Pino}
	\thanks{R. D. and J. d. P. contributed equally to this work.}
	\affiliation{Center for Nanophotonics, AMOLF, Science Park 104, 1098 XG Amsterdam, The Netherlands}
	\author{Ewold Verhagen}
	\affiliation{Center for Nanophotonics, AMOLF, Science Park 104, 1098 XG Amsterdam, The Netherlands}
	\author{Andrea Alù}
	\affiliation{Department of Electrical and Computer Engineering, The University of Texas at Austin, Austin, TX 78712, USA}
	\affiliation{Photonics Initiative, Advanced Science Research Center, City University of New York, NY 10031, USA}
	\affiliation{Physics Program, Graduate Center, City University of New York, NY 10016, USA }
	\affiliation{Department of Electrical Engineering, City College of The City University of New York, NY 10031, USA}
	\email{aalu@gc.cuny.edu}
	
	\begin{abstract}
		We demonstrate an optomechanical platform where optical mode conversion mediated by mechanical motion enables arbitrary tailoring of polarization states of propagating light fields. Optomechanical interactions are realized in a Fabry-Perót resonator, which naturally supports two polarization-degenerate states while an optical control field induces rotational symmetry breaking. Applying such principles, the entire Poincaré sphere is spanned by just optical control of the driving field, realizing reciprocal and non-reciprocal optomechanically-induced birefringence for linearly polarized and circularly polarized control driving. A straightforward extension of this setup also enables all-optical tunable isolation and circulation. Our findings open new avenues to exploit optomechanics for arbitrary manipulation of light polarization.
	\end{abstract}
	
	
	\maketitle
	
	Generation of arbitrary light polarization states is of major fundamental and technological relevance in nonlinear and quantum optics~\cite{Nielsen2010}, communication networks~\cite{xiong2018complete} and microscopy~\cite{Inoue1981,DeBoer1997}. Reciprocal and non-reciprocal polarization control is conventionally attained in bulk optics by wave plates and Faraday rotators, respectively, which rely on anisotropic or magnetically biased materials to induce birefringence. Both effects are intrinsically weak in bulk materials, requiring sizable driving fields and structures larger than the wavelength. More recently, engineered metasurfaces~\cite{Zhao2011,Pfeiffer2016,Grady2013,Arbabi2015} have been developed to overcome these limitations across the electromagnetic spectrum, with tunability enabled by both electric~\cite{Wu2019} and magnetic~\cite{Floess2015,Fallahi2012} biasing. Polarization rotation is nevertheless limited, and no platform offers complete tunability of the polarization state in a compact structure. 
	
	Optomechanical technologies exploit sharp photonic resonances coupled to mechanical modes to enable a new degree of control over light, examples of which are the swapping of photons into phonons and back~\cite{Safavi-Naeini2011,Fiore2011, Verhagen2012,Palomaki2013} and optomechanically induced transparency~\cite{Weis2010, Safavi-Naeini2011b}. In this context, multimode systems have been recognized for their potential to convert photons from one cavity mode to another one via mechanical excitations. This additional degree of control over light allows quantum and classical applications, such as wavelength conversion between two distant optical frequencies ~\cite{Regal2011,Dong2012,Hill2012,Andrews2014} that can feature adiabatic quantum state transfer~\cite{Tian2010,Tian2012}, non-demolition measurements~\cite{Lee2015},  entanglement generation~\cite{Wang2013,Wang2015,Kuzyk2013}, non-reciprocal transport and optical routing~\cite{Shen2016,Ruesink2016,Fang2017,Peterson2017,Ruesink2018}. For the large part, these proposals have been so far based on manipulating the scalar properties of photons, while their vector nature has only been recently pinpointed~\cite{Xiong2016,Buters2016,zanotto2018optomechanics}. 
	
	\begin{figure}
		\centering
		\includegraphics[width=\columnwidth]{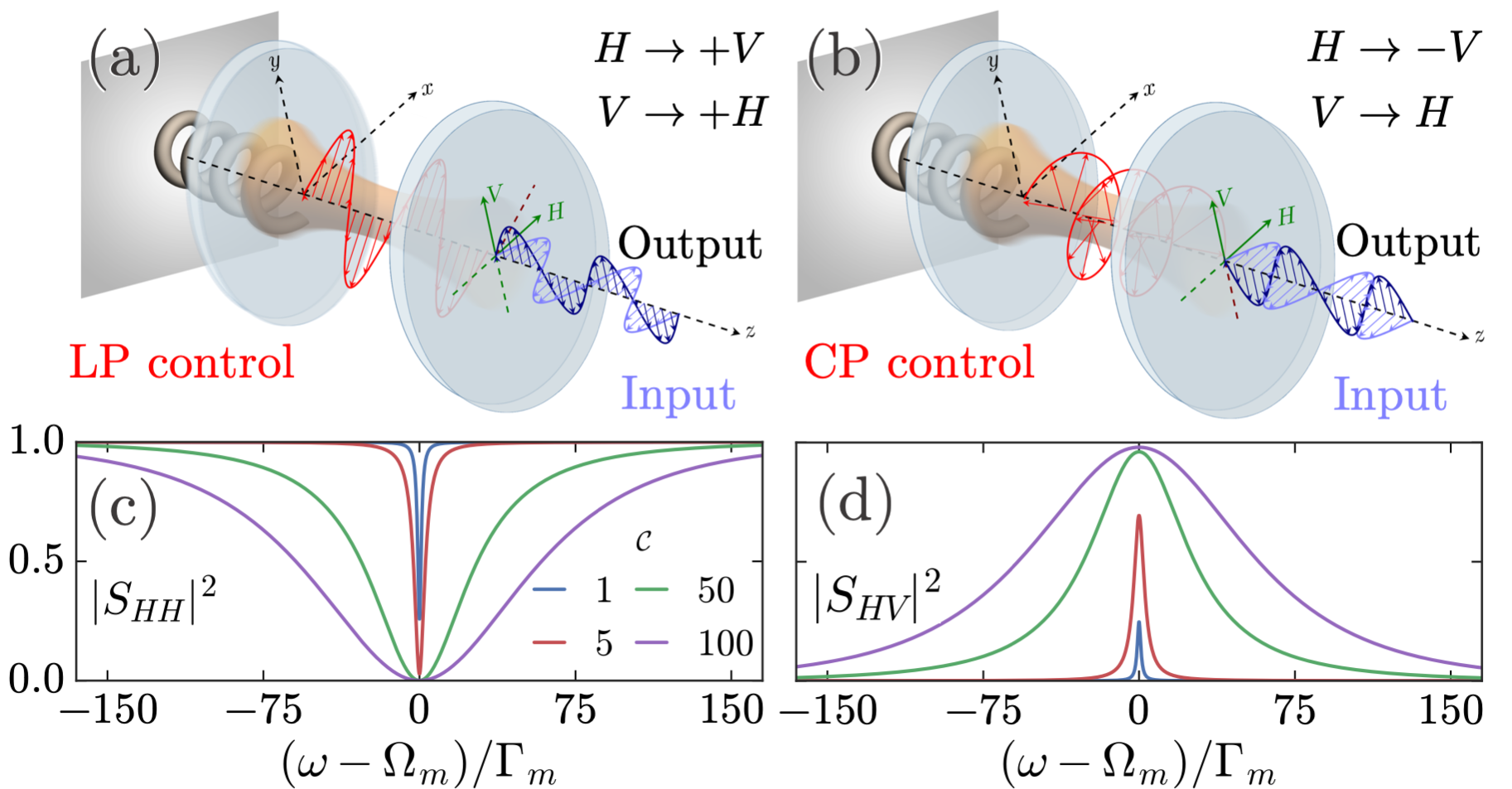}
		\caption{(a,b) Sketch of the proposed geometry. A pump field (red) induces coupling between two degenerate cavity modes with orthogonal polarizations, altering the polarization state of a probe (blue) upon reflection (black). Lower panels show the reflectivity spectra for $H$-polarized input into the $H$ and $V$ output channels. The output is reciprocal ($\mathcal{S}_{HV}=\mathcal{S}_{VH}$) for a linearly-polarized pump, or non-reciprocal for a circularly-polarized pump ($\mathcal{S}_{HV}=-\mathcal{S}_{VH}$),  inducing optical activity (birefringence) and Faraday rotation, respectively. The amplitudes of the co- and cross-polarized reflected fields are the same for both pumps, so (c,d) apply to both (a,b). For these and later plots, $\Omega_{m}=2\pi \times50$ GHz, $\Gamma_{m}=2\pi \times10$MHz, $\kappa=\kappa_{c}=2\pi\times5$ GHz.}
		\label{fig1}
	\end{figure}
	
	In this Letter, we demonstrate how optomechanically induced mode conversion enables arbitrary control of optical polarization. For simplicity and clarity of presentation, we focus on a canonical optomechanical system, consisting of a high-$Q$ planar Fabry-Pérot resonator with a movable mirror, as is used in many experiments (see  \autoref{fig1}) ~\cite{abramovici1992ligo,Cohadon1999,Corbitt2006,Thompson2008,Favero2009,Groblacher2009,Purdy2012,Buters2016,Anguiano2017,Rossi2018,kharel2018high}.  This system naturally hosts degenerate modes of orthogonal polarization that are both coupled to a mechanical resonator mode, while similar concepts, however, may be envisioned in a broad class of 2D and 3D geometries. Optical pumping in a particular polarization state breaks rotational symmetry and induces a tunable linear interaction between different polarizations. In contrast to static radiation-pressure-induced chirality in pre-patterned metasurfaces~\cite{Liu2017}, our approach exploits dynamical back-action effects and coherent mode conversion for fully tunable and efficient polarization control. Indeed, by bringing ideas of coherent polarization manipulation via electromagnetically-induced transparency in atomic systems~\cite{Wielandy1998} into the realm of linear optomechanics, we allow a rich regime of reciprocal and non-reciprocal interactions, within an on-chip scalable platform that enables tunable all-optical isolation and circulation.

	Our model includes two degenerate electromagnetic modes with equal frequencies $\omega_\mathrm{cav}$, decay rates $ \kappa$ and annihilation operators $\hat{\textbf{a}}^T=(\hat{a}_{H},\hat{a}_V)$ ($H,V$ label horizontal and vertical polarization axes), interacting with a localized mechanical mode, with frequency $\Omega_m$, damping constant $\Gamma_m$, and annihilation operator $\hat{b}$. We first consider the end-coupled geometry shown in \autoref{fig1}, consisting of a highly-reflecting movable mirror that supports a mechanical mode and a partially transparent input/output port. The resonator, assumed to satisfy the resolved sideband condition ($\kappa\ll\Omega_m$), is driven by a red-detuned control field, characterized by the steady-state Jones vector $\bar{\boldsymbol{\alpha}}^T=(\bar{\alpha}_{H},\bar{\alpha}_V)$ at a frequency  $\omega_L=\tilde{\Delta}+\tilde{\omega}_\mathrm{cav}$. Here $\tilde{\omega}_\mathrm{cav}$ includes a static blue shift proportional to the average radiation pressure force $\bar{F}\propto g_0|\bar{\boldsymbol{\alpha}}|^2$.  In a frame rotating at red-detuned frequency $\omega_L$ close to resonance ($\tilde{\Delta}=-\Omega_m$), the photons associated with the `probe' fields in side bands of the stronger control field, denoted by  $\delta\hat{\textbf{a}}\simeq\hat{\textbf{a}}-\bar{\boldsymbol{\alpha}}$, excite maximally the mechanical mode. Coherent dynamics is thus governed by the linearized Hamiltonian within the rotating-wave approximation (setting $\hbar=1$),
	\begin{equation}
		\hat{H}_\mathrm{eff}=-\tilde{\Delta}\delta\hat{\textbf{a}}^{\dagger}\delta\hat{\textbf{a}}+\Omega_m\hat{b}^\dagger\hat{b}-g_0(\bar{\boldsymbol{\alpha}}\cdot\delta\hat{\textbf{a}}^{\dagger}\hat{b}+\mathrm{H.c.})\label{eq:Heffx},
	\end{equation}
	where $g_0$ denotes the vacuum optomechanical coupling. In a way similar to conversion between photonic modes with different wavelengths~\cite{Dong2012,Hill2012,Andrews2014}, the synthetic optomechanical interaction in \autoref{eq:Heffx} implements a complex parametric coupling between polarization-orthogonal photons that are degenerate. Essentially, a photon in mode $H$ can be annihilated to produce a phonon in the mechanical resonator, which can subsequently be annihilated to produce a photon in mode $V$ (and vice versa). As we demonstrate below, this enables full spanning of the Poincaré sphere (PS)  with the output reflected from a single resonator.
	
	To characterize our system we obtain the semiclassical equations of motion derived from the coherent evolution \autoref{eq:Heffx} after including dissipation, which for the photonic modes are split into in-coupling and intrinsic loss rates, $\{\kappa_c,\kappa_i\}$, ($\kappa=\kappa_c+\kappa_i$). The solution for $\hat{b}(\omega)$ is readily found in Fourier space and reinserted into the photonic evolution, yielding the linear system $i\omega\delta\hat{\textbf{a}}(\omega)=i\mathcal{M}(\bar{\boldsymbol{\alpha}})\delta\hat{\textbf{a}}(\omega)+\sqrt{\kappa_c}\textbf{s}_\mathrm{in}$, where $\textbf{s}_\mathrm{in}$ contains probe modes. The $2\times2$ scattering matrix linking input to output modes ($\textbf{s}_\mathrm{out}=\mathcal{S}\textbf{s}_\mathrm{in}$) reads ~\cite{Miri2017}
	\begin{equation}
		\mathcal{S}(\omega,\bar{\boldsymbol{\alpha}})=-\textbf{1}+i\kappa_c(\mathcal{M}(\bar{\boldsymbol{\alpha}})+\omega\textbf{1})^{-1},\label{eq:S_matrix}
	\end{equation}
	where direct reflection is included via the first term and $\textbf{1}=\mathrm{diag}(1,1)$. In the following we employ the cooperativity $\mathcal{C}=4 g_0^{2}|\bar{\boldsymbol{\alpha}}|^{2}/(\Gamma_{m}\kappa)$ as a relevant figure of merit for the optomechanical system. Further details of this derivation and the resulting $\mathcal{S}$-matrix can be found in the Supplemental Material ~\cite{sm}.    
	
	\begin{figure*}
		\centering
		\includegraphics[width=\linewidth]{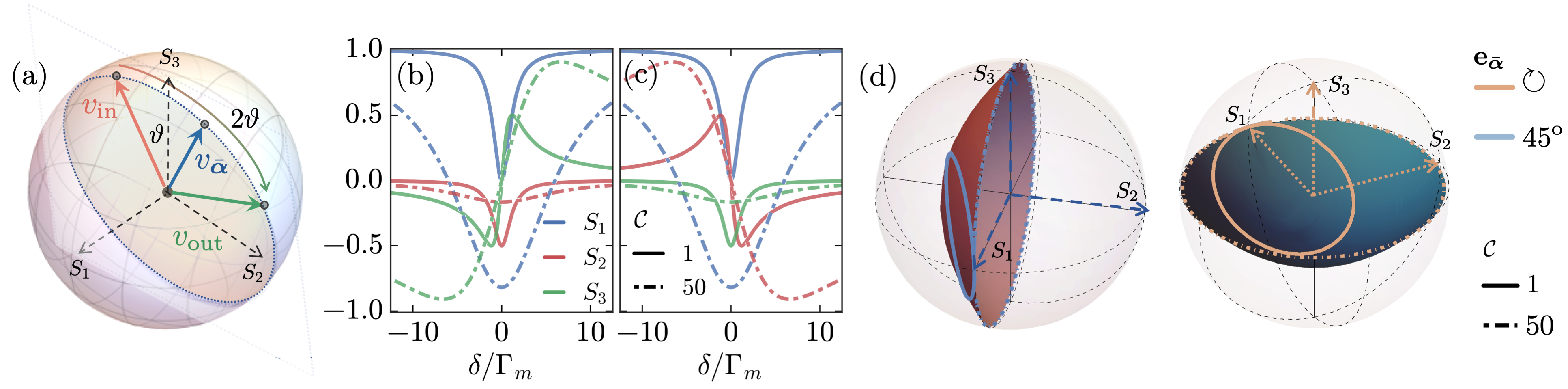}
		\caption{(a) Illustration of the geometrical transformation induced in the PS by the scattering matrix in \autoref{eq:S_matrix}. (b,c,d):
			Stokes parameters for $H$-polarized incident light and varying input frequency with $\mathcal{C}=1$ (dashed) and  $\mathcal{C}=50$ (solid) when the pump is linearly polarized at 45º (b) or right-handed circularly polarized (c).  The surface of the sphere can be reached in the limit $\mathcal{C}\gg 1$, but even at moderate levels shown here, the output state can nearly be in 5 of the 6 Stokes basis states, at the expense of photon losses. (d) Manifolds showing output states for varying cooperativity and detuning ($\delta\in(-\infty,\infty)$) for a linearly polarized pump along 45º (left) and right-handed circular polarization (right). The paths inside the PS for fixed $\mathcal{C}=\{1,50\}$ are mapped out, with different pump power and polarization, corresponding to (b) and (c). For panels (b),(c),(d), other parameters are chosen according to ~\autoref{fig1}.}
		\label{fig2}
	\end{figure*}
	
	We explore first the case of linear incident polarization with a finite angle with respect to a linearly polarized control field.  The reflectivity spectra for the co- and cross-polarized components of $\textbf{s}_\mathrm{in}\parallel\textbf{e}_H$, given by $\mathcal{S}_{HH}$ and $\mathcal{S}_{VH}$ are displayed in \autoref{fig1}c,d for the case $\bar{\boldsymbol{\alpha}}\parallel\textbf{e}_{45º}=(\textbf{e}_H+\textbf{e}_V)/\sqrt{2}$.  The control field thus explicitly breaks rotational symmetry and produces an optomechanically induced mixture of otherwise uncoupled orthogonal polarizations, channeled by a narrowband mechanical resonance (with effective linewidth $(C+1)\Gamma_\mathrm{m}$).  This phenomenon, emulating a power-controlled birefringence, is characterized by the resonant scattering-parameters which only depend on the mechanics via $\mathcal{C}$,
	\begin{subequations}\label{eq:conv_linear}
		\begin{align}
			\mathcal{S}_{H H}(\Omega_m,\bar{\boldsymbol{\alpha}}\parallel\textbf{e}_{45º})=&\frac{\kappa_{c}-(\mathcal{C}+1)\kappa_i}{(\kappa_{c}+\kappa_i)(\mathcal{C}+1)},\\
			\mathcal{S}_{VH}(\Omega_m,\bar{\boldsymbol{\alpha}}\parallel\textbf{e}_{45º})=&\frac{\mathcal{C}\kappa_{c}}{(\kappa_{c}+\kappa_i)(\mathcal{C}+1)}.
		\end{align}    
	\end{subequations}
	With increasing driving intensity, the peak and the bandwidth of the conversion are augmented. This efficiency saturates in the large cooperativity limit ($\mathcal{C}\gg 1$) at $|\mathcal{S}_{VH}|^2\simeq\kappa_c^2/(\kappa_c+\kappa_i)^2$ and becomes unitary ($|\mathcal{S}_{HH}|^2=0,|\mathcal{S}_{VH}|^2=1$) for negligible optical losses $\kappa_i\ll\kappa_c$. A critical aspect in maximizing the conversion efficiency of this process at larger powers is the fulfillment of the sideband-resolved condition, which for fixed detuning ultimately limits the bandwidth via $\tilde{\omega}_\mathrm{cav}-\omega_\mathrm{cav} <\kappa$. The cavity linewidth itself also imposes the bound $\left|\omega-\tilde{\omega}_\mathrm{cav}\right| <\kappa$.  
	
	As revealed by the interaction Hamiltonian in \autoref{eq:Heffx}, the mechanical mode selectively couples with the superposition of photonic modes $\bar{\boldsymbol{\alpha}}\cdot\delta\hat{\textbf{a}}$. The conversion into orthogonal modes $\sim\bar{\boldsymbol{\alpha}}'\delta\hat{\textbf{a}}$, where $\langle\bar{\boldsymbol{\alpha}},\bar{\boldsymbol{\alpha}}'\rangle=0$, is then unaffected by optomechanics and only a shift in the cavity resonance is expected. For the same control field assumption as above, the scattering matrix elements for parallel and orthogonal conversion channels,
	\begin{subequations}\label{eq:S_45}
		\begin{align}
			\mathcal{S}_{{45º} H}(\Omega_m,\bar{\boldsymbol{\alpha}}\parallel\textbf{e}_{45º})=&-\frac{(\mathcal{C}-1)\kappa_{c}+(\mathcal{C}+1)\kappa_i}{\sqrt{2}(\kappa_{c}+\kappa_i)(\mathcal{C}+1)},\\  
			\mathcal{S}_{{-45º} H}(\Omega_m,\bar{\boldsymbol{\alpha}}\parallel\textbf{e}_{45º})=&\frac{\kappa_{c}-\kappa_i}{\sqrt{2}(\kappa_{c}+\kappa_i)}\label{eq:S_circD},    
		\end{align}
	\end{subequations}
	demonstrate this exact phenomena. Only the probe component co-polarized with the pump exhibits optomechanically induced transparency or absorption, depending on $\kappa_i$, recognized in~\cite{Xiong2016}. In the limit of low loss and high $\mathcal{C}$, this component undergoes a $\pi$ phase shift, resulting in strong polarization conversion. 
	
	In our system, conversion of polarization states depends not only on the control intensities, determining the absolute magnitude of the coupling constant in \autoref{eq:Heffx}, but are also controlled by the pump phases $\arg(\bar{\alpha}_{H, V})$. The potential of this phase manipulation is uncovered by considering the case of right-handed and left-handed circular polarizations (RHCP, LHCP), given by $\{\textbf{e}_\circlearrowright=(\textbf{e}_H+i\textbf{e}_V)/\sqrt{2},\textbf{e}^*_\circlearrowleft=\textbf{e}_\circlearrowright\}$ independent of the propagation direction to avoid confusion upon reflection. With a fixed gauge $\mathrm{arg}(\bar{\alpha}_H\bar{\alpha}_V^*)=-\pi/2$, it is straightforward to check that the conversion efficiencies of the $H$-input probe into the circular bases are equivalent to the the conversion into the $\pm45º$ bases under linear pumps (\autoref{eq:S_45}), i.e. $\mathcal{S}_{\circlearrowright H (\circlearrowleft H)} (\bar{\boldsymbol{\alpha}}\parallel\bold{e_{\circlearrowright}})=\mathcal{S}_{45º H (-45º H)} (\bar{\boldsymbol{\alpha}}\parallel\bold{e}_{45º})$. 
	
	Hence, by tailoring the static radiation pressure in a Fabry-Perót cavity, optomechanical interactions permit to leverage independently amplitude and phase of the four $\mathcal{S}$-matrix elements, enabling arbitrary polarization control. The relations fulfilled by conversion efficiencies for parallel/orthogonal modes with respect to control field suggest a geometrical interpretation for the parametric action of the $\mathcal{S}$-matrix, as we develop in the following. At large cooperativities,  
    a basis-independent expression for the resonant $\mathcal{S}$-matrix shows the conversion is insensitive to the mechanical degree of freedom (encoded in $\Gamma_m,g_0$), namely $\mathcal{S}(\Omega_{m},\bar{\boldsymbol{\alpha}})\mathbf{s}_\mathrm{in}\simeq-\mathbf{s}_\mathrm{in}+2\kappa_{c}\left(\mathbf{s}_\mathrm{in}-\langle\textbf{e}_{\bar{\boldsymbol{\alpha}}},\mathbf{s}_\mathrm{in}\rangle\textbf{e}_{\bar{\boldsymbol{\alpha}}}\right)/\kappa$, with the control polarization vector $\textbf{e}_{\bar{\boldsymbol{\alpha}}}=\bar{\boldsymbol{\alpha}}/|\bar{\boldsymbol{\alpha}}|$, playing the role of a (complex) reflection axis for $\textbf{s}_\mathrm{in}$ (See ~\cite{sm} for further details). To gain a deeper insight, we use the natural representation of polarization states in the Poincaré sphere that displays the Stokes parameters $(S_0,S_1,S_2,S_3)$, representing the degree of polarization along the bases $\{\textbf{e}_H$, $\textbf{e}_V\}, \{\textbf{e}_{+45},\textbf{e}_{-45}\}$, and $\{\textbf{e}_\circlearrowleft,\textbf{e}_\circlearrowright\}$~\cite{collett2005field}. The input Jones vector $\textbf{s}_\mathrm{in}=S_0^\mathrm{in}\left(\cos(\theta_\mathrm{in}/2),e^{i\varphi_\mathrm{in}}\sin(\theta_\mathrm{in}/2)\right)$ is thus replaced by the Stokes 3-vector $\textbf{v}_{\mathrm{in}}=S_{0}^\mathrm{in}\left(\cos(\theta_{\mathrm{in}})\sin(\varphi_{\mathrm{in}}),\sin(\theta_{\mathrm{in}})\sin(\varphi_{\mathrm{in}}),\cos(\varphi_{\mathrm{in}})\right)$, while the $\mathcal{S}$-mapping exactly induces a reflection of the input vector over the parametric plane containing the control vector $\textbf{v}_{\bar{\boldsymbol{\alpha}}}$ for an overcoupled resonator ($\kappa_i\ll\kappa_c$), namely  $\textbf{v}_\mathrm{out}\simeq2(\textbf{v}_{\bar{\boldsymbol{\alpha}}}\cdot\mathbf{v}_\mathrm{in})\textbf{v}_{\bar{\boldsymbol{\alpha}}}-\mathbf{v}_\mathrm{in}$   ~\cite{Ranzani2014}. The whole surface of the PS can therefore be spanned by rotating $\textbf{e}_{\bar{\boldsymbol{\alpha}}}$ (see \autoref{fig2}a).
	
	\begin{figure*}
		\centering
		\includegraphics[width=\linewidth]{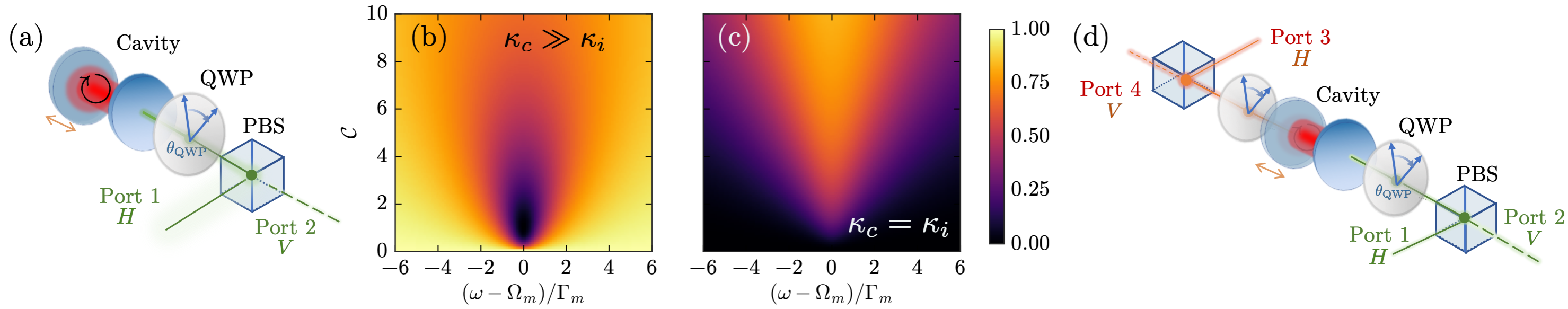} \caption{
			(a) Schematic for the realization of a free-space optical isolator. (b) Transmission $|\mathcal{S}_{VH}|^2$ for the over-coupled case, where $|\mathcal{S}_{HV}|\simeq 1$. Good isolation occurs for $\mathcal{C}\simeq 1$ and over the mechanical linewidth, where losses through the mechanical mode can occur. (c) Transmission $|\mathcal{S}_{VH}|^2$ for the critically coupled case. With larger pumping the transmission amplitude and bandwidth both increase. (d) Schematic for a potential implementation as a 4-port circulator. For panels (b),(c), other parameters are chosen according to \autoref{fig1}.}
		\label{fig3}
	\end{figure*}

	The  effects studied so far to attain polarization control would require active control of the driving polarization. To outline these constrains, we next explore how driving frequencies can be exploited as an additional degree of freedom in the conversion process. For instance, when the probe frequency matches the optical resonance, a finite detuning of the control field (denoted by $\delta=\Omega_m-\omega_L$) causes the output field to acquire a different phase. To see to which extent the output state is thus modified, we show in \autoref{fig2}b the Stokes parameters for the reflected state of a cavity driven by $H$-polarized light, with an LP control field along 45º. Recovering previous results, we observe full dissipation of the co-polarized probe component into the mechanical mode at resonance ($\delta=0$) for an overcoupled resonator ($\kappa_i\ll\kappa_c$) driven at moderate cooperativities  ($\mathcal{C}\sim1$). The system then behaves as a conventional polarizing mirror, reflecting only light cross-polarized to the control. Similar physics is observed for an RHCP control in \autoref{fig2}c,  where reflection is purely LHCP ($S_3<0$). Efficient conversion into the polarization state $\textbf{e}_V$, sitting at $S_1=-1$, is recovered at $\mathcal{C}\gg1$ for both cases. Out of resonance ($\delta\neq0$),  signatures of conversion into elliptic polarization ($S_{1,2,3}\neq0$) are noticed with an output state increasingly approaching the surface of the PS for large cooperativity. 
	
	Furthermore, within the geometrical argument presented in \autoref{fig2}a, finite detuning results in output Stokes vectors with a component orthogonal to the control plane leading to extra control over output polarization beyond resonant conversion.  In order to benchmark this approach, we trace in \autoref{fig2}d the manifold of reachable Stokes parameters on the PS as $\delta$ is swept across the range $\delta\in(-\infty,\infty)$ and the cooperativity is increased above zero, i.e. $\mathcal{C}>0$, for the $45º$ and RHCP control polarizations (left/right panels). The loci of output states form bowl-shaped surfaces that emerge from the zero-conversion point at $S_1=1$ (state $\textbf{e}_H$) for $\mathcal{C}=0$,  and infiltrate the PS for intermediate $\mathcal{C}$. The output reaches the state $\textbf{e}_V$ at the (antipodal) location $S_1=-1$, obtained by reflecting the input across the parametric control plane in the limit $\mathcal{C}\gg1$ and resonance ($\delta=0$), recovering previous results. For different control polarizations, these surfaces show inclination planes normal to $\textbf{v}_{\bar{\boldsymbol{\alpha}}}$ and therefore orthogonal to each other, provided $\textbf{v}_{45º}\cdot\textbf{v}_\circlearrowright=0$. From a global perspective, the control field Stokes vector serves as an axis of rotation for an aribitrary probe field as the frequency is varied, with a maximum $180º$ rotation for $\delta=0$, as in \autoref{fig2}a. While output states are identical in the large cooperativity limit for linear and circular polarizations, the additional phase accumulated for finite detuning enables access to different regions of the PS. In particular, at fixed cooperativity, output states follow circular orbits with radii growing with $\mathcal{C}$. In this fashion, it is feasible to reach four out of the six poles of the PS along the axes $S_{1,2,3}$ by just varying detuning. An extra 5th pole of the PS is reached at the expense of losses in the polarizer operation. This demonstrates the potential of this system to devise a reconfigurable polarization system. 
	
	Interestingly, finite ellipticity of the control field induces conversion processes that are non-reciprocal in nature. In the case of circular polarization, the scattering matrix is antisymmetric, signaling the breaking of Lorentz reciprocity ($\mathcal{S}=-\mathcal{S}^T\neq \mathcal{S}^T$, see Supplemental Material ~\cite{sm}).  Antisymmetry of the scattering matrix is inherited from the complex phase $\mathrm{arg}(\bar{\alpha}_H^*\bar{\alpha}_V)$ imprinted by the control field at the level of the interaction Hamiltonian in \autoref{eq:Heffx}.  
	Such features show time-reversal symmetry breaking and (non-reciprocal) Faraday rotation in the system and can be noticed in the PS as the inability to relate forward/backward scattering transformations~\cite{Ranzani2014}. 
	
    Phase non-reciprocity can be readily exploited to build devices displaying unconventional photon routing~\cite{Verhagen2017}, such as an optical isolator, where the off-diagonal transmission coefficients have unequal amplitudes. A possible approach consists in placing a suitably oriented quarter-wave plate (QWP) after the output mirror of a CP-driven resonator (without loss of generality RHCP) in addition to a polarizing beam splitter (PBS), forming the ports sketched in \autoref{fig3}a. Then, an $H$-polarized input probe (port 1) is reflected in the PBS and acquires an RHCP component upon passing the QWP. This component yields a $V$-polarized output (port 2), after a second pass through the QWP, with a transmission level tunable by interactions with the mechanical mode. Following a similar logic, a $V$-polarized probe in port 2 is insensitive to the mechanics and has a fixed transmission to port 1. In this system losses, inherently required by a linear two-port isolator ~\cite{Pozar2012}, are found in preferential absorption of one of the CP components.
	
	This is confirmed by quantitative analysis of the device action on the input, characterized by $\mathcal{S}_\mathrm{iso}=\mathcal{A}_\mathrm{QWP}\mathcal{S\mathcal{A}_\mathrm{QWP}}$ where $\mathcal{A}_\mathrm{QWP}=\frac{1}{\sqrt{2}}\left(\mathbf{1}+i\sigma_{x}\right)$ and $\mathcal{S}$ follows from \autoref{eq:S_matrix}. For illustrative purposes, we consider two limiting scenarios, in which the optical cavity is either \textit{i)} over-coupled ($\kappa_i\ll\kappa_c$), or \textit{ii)} critically coupled ($\kappa_i=\kappa_c$)~\cite{Miri2017}. If the control field in the resonator is set to RHCP, then in the former case, the RHCP probe can be absorbed through the mechanical modes with moderate cooperativities, leading to vanishing reflected output, i.e., $\mathcal{S}_{VH}\simeq0$ (see \autoref{fig3}b). Meanwhile, the LCHP light will not interact with the mechanical mode and will be reflected independent of cooperativity ($|\mathcal{S}_{HV}|\simeq1$). Efficient isolation occurs for $\mathcal{C}=1$ over a narrow bandwidth limited by the mechanical resonance, $~\Gamma_m$. For critical coupling, light is absorbed without interacting with the mechanical mode ($\mathcal{S}\simeq \textbf{0}$ for low $\mathcal{C}$ in \autoref{fig3}c). In this case, the $\mathcal{S}_{VH}$ signal exhibits features reminiscent from optomechanically-induced transparency, with increasing efficiency and bandwidth for large cooperativities, similar to polarization conversion in the overcoupled cavity displayed in \autoref{fig1}d. Similarly, $\mathcal{S}_{HV}\simeq0$ regardless of pump power. Arbitrary contributions of $\kappa_c$ and $\kappa_i$ yield a behavior interpolating between the two limits above. 
	
	If the reflectivity of the second mirror is decreased below unity, so light entering in the resonator can out-couple to the other side, the addition of the same QWP-PBS plate to this side, as sketched in \autoref{fig3}c, permits non-reciprocal light circulation between a port $i$ and a port $i+1$, for $i=\{1,2,3,4\}$. This setup is precisely mappable into an optomechanically-assisted four-port optical circulator~\cite{Ruesink2016}. A benchmark of this device can be found in the Supplemental Material~\cite{sm}. 
	
	To conclude, we have demonstrated how a minimal setup, consisting of an optomechanical cavity in the resolved-sideband regime, is a versatile platform for all-optical polarization conversion. Parametric photon-phonon interactions induce birefringence for a probe beam in a rotationally-symmetric system. Our results then highlight how the freedom in both vectorial (polarization) and scalar (intensity and frequency) degrees of freedom in the incoming fields can be exploited to achieve arbitrary polarization states in reflection.  In particular, we have shown how birefringence is expected for linearly polarized control beams, and similarly, non-reciprocal Faraday rotation can be realized for circular polarization. Finally, the design of an optical isolator based upon this configuration is presented,  showing the requirements and tradeoffs that would exist in a realistic implementation of such a system. These concepts may be straightforwardly extended to other settings supporting degenerate optomechanical resonances, such as integrated photonics platforms, or by exploiting stimulated Brillouin scattering in polarization-degenerate waveguides or fibers~\cite{VanDeventer1994,Aryanfar2014}.

	Moreover, optomechanical amplification,  which is achieved by blue-detuned control fields at $\tilde{\Delta}= \Omega_m$, may also open novel scenarios where the resulting parametric gain may relax cooperativity demands for efficient birefringence, or create directional amplification in a setup similar to our isolator. Our findings thus unlock a new potential of optomechanics in manipulating light fields by interfacing with mechanical degrees of freedom, enabling unusual reciprocal and non-reciprocal phenomena for polarization conversion and manipulation.
	
	Acknowledgements: E.V. thanks Frank Buters, whose thesis defense inspired thinking about the proposed concepts. The authors acknowledge support from the Office of Naval Research (grant no. N00014-16-1-2466) and the Air Force Office of Scientific Research. This work is part of the research programme of the Netherlands Organisation for Scientific Research (NWO). It is furthermore supported by the European Research Council (ERC Starting Grant no. 759644-TOPP) and the European Union's Horizon 2020 research and innovation programme under grant agreement no. 732894 (FET Proactive HOT).

	\bibliography{OM_BFE} 
	
\end{document}


\title{Supplemental Material: Optomechanically Induced Birefringence and Faraday Effect}
	
	\author{Robert Duggan}
	\affiliation{Department of Electrical and Computer Engineering, The University of Texas at Austin, Austin, TX 78712, USA}
	\author{Javier del Pino}
	\affiliation{Center for Nanophotonics, AMOLF, Science Park 104, 1098 XG Amsterdam, The Netherlands}
	\author{Ewold Verhagen}
	\affiliation{Center for Nanophotonics, AMOLF, Science Park 104, 1098 XG Amsterdam, The Netherlands}
	\author{Andrea Alù}
	\affiliation{Department of Electrical and Computer Engineering, The University of Texas at Austin, Austin, TX 78712, USA}
	\affiliation{Photonics Initiative, Advanced Science Research Center, City University of New York, NY 10031, USA}
	\affiliation{Physics Program, Graduate Center, City University of New York, NY 10016, USA }
	\affiliation{Department of Electrical Engineering, City College of The City University of New York, NY 10031, USA}

	\maketitle
	
		\section{Semiclassical scattering analysis}
		Here we detail the derivation of the scattering matrix of the system. The starting point is the standard optomechanical Hamiltonian for two degenerate cavity modes $\hat{\textbf{a}}^T=(\hat{a}_H,\hat{a}_V)$ in the rotating frame of a biasing pump field at frequency $\omega_L$, \cite{bowen2015quantum}, namely (setting $\hbar=1$)
		\begin{equation}
			\hat{H}=\hat{H}_\mathrm{mec}-\hat{\textbf{a}}^{\dagger}\hat{\textbf{a}}\big(\Delta + G\hat{x}\big),\label{eq:full_H}
		\end{equation}
		where  $\hat{H}_\mathrm{mec}=\frac{m_\mathrm{eff}\Omega_m}{2}\hat{x}^2+\frac{\hat{p}^2}{2m_\mathrm{eff}}$ for an oscillator with effective mass $m_\mathrm{eff}$.  The last term in \autoref{eq:full_H} introduces a strong biasing pump via $\hat{H}_{\mathrm{in}}\sim\bar{\boldsymbol{\alpha}}^*\cdot\hat{\textbf{a}}+\mathrm{H.c.}$. Linearization around the steady state amplitude $\bar{\boldsymbol{\alpha}}$ yields
		\begin{equation}
			\hat{H}_\mathrm{eff}=\hat{H}_\mathrm{mec}-\tilde{\Delta}\delta\hat{\textbf{a}}^{\dagger}\delta\hat{\textbf{a}}-G(\bar{\boldsymbol{\alpha}}\cdot\delta\hat{\textbf{a}}^{\dagger}+\boldsymbol{\alpha}^*\cdot \delta\hat{\textbf{a}})\hat{x}, \label{eq:linear_H}
		\end{equation}
		defined the effective detuning $\tilde{\Delta}=\Delta+2g_0^2|\bar{\boldsymbol{\alpha}}|^2/\Omega_m$. Neglecting mechanical noise, the system evolution under a weak probe $\textbf{s}_\mathrm{in}$ follows from the equations of motion: 
		\begin{subequations}
         \label{eq:time_evo1}
			\begin{align}
			\tag{S3a}\label{eq:time_evo1a}
				\dot{\delta\hat{\boldsymbol{a}}}=&i\left(\tilde{\Delta}-\frac{\kappa}{2}\right)\delta\hat{\textbf{a}}+iG\bar{\boldsymbol{\alpha}} \hat{x}+\sqrt{\kappa_c}\textbf{s}_\mathrm{in},\\
				\tag{S3b}\label{eq:time_evo1b}
				\ddot{\hat{x}}=&-\Omega_{m}^{2} \hat{x}-\Gamma_{m}\dot{\hat{x}}+\frac{G}{m_\mathrm{eff}}\left(\bar{\boldsymbol{\alpha}}^{*}\cdot\delta \hat{\textbf{a}}+\mathrm{H.c.}\right). 
			\end{align}
		\end{subequations}
	    From \autoref{eq:time_evo1}, the mechanical oscillations act as sources for the optical modes, with strengths and phases directly related to the steady-state biasing fields. The same holds for the optical forces. Solving in Fourier space for the long-time mechanical amplitudes in  \autoref{eq:time_evo1b}, \autoref{eq:time_evo1a} reads as $-i\omega\delta\hat{\boldsymbol{a}}(\omega)=i\mathcal{M}(\hat{\boldsymbol{\alpha}})\delta\hat{\boldsymbol{a}}(\omega)+\textbf{s}_\mathrm{in}$ with
	    \begin{equation}
	    i(\mathcal{M}\left(\bar{\boldsymbol{\alpha}}\right)+\textbf{1}\omega)=\chi_{a}^{-1}(\omega)\left[\mathbf{1}-G^{2}\chi_{a}(\omega)\chi_{m}(\omega)\left(\bar{\boldsymbol{\alpha}}\bar{\boldsymbol{\alpha}}^{\dagger}\right)\right],
	    \end{equation}
	    where  $\chi_{m}(\omega)=(m_\mathrm{eff}(\Omega_m^2-\omega^2)-im_\mathrm{eff}\Gamma_m\omega)^{-1}$ and  $\chi_{\alpha}(\omega)=(-i(\Delta+\omega)+\kappa/2)^{-1}$ stand for the mechanical and cavity susceptibilities, respectively. Input and output modes are linked by the $2\times2$ scattering matrix (basis-independent) expression $  \mathcal{S}(\omega,\bar{\boldsymbol{\alpha}})=-\textbf{1}+i\kappa_c(\mathcal{M}(\bar{\boldsymbol{\alpha}})+\omega\textbf{1})^{-1}$.  Since $\mathcal{M}(\bar{\boldsymbol{\alpha}})$ can be written as the outer product of two vectors, the inverse follows from the following corollary of the Woodbury matrix identity \cite{Bartlett1951}: for general $\mathbf{u},\mathbf{v}, (\mathbf{1+}\mathbf{u}\mathbf{v}^{T})^{-1}=\mathbf{1-\mathbf{u}\mathbf{v}^{T}/(1+\mathbf{v}^{T}\mathbf{u}})$, hence
	    \begin{equation}
	    \mathcal{S}=-\textbf{1}+\kappa_c\chi_{a}(\omega)\left[\mathbf{1}-\frac{G^{2}\chi_{a}(\omega)\chi_{m}(\omega)\bar{\boldsymbol{\alpha}}\bar{\boldsymbol{\alpha}}^{\dagger}}{1+G^{2}|\bar{\boldsymbol{\alpha}}|^{2}\chi_{a}(\omega)\chi_{m}(\omega)}\right]\hspace{-4mm}.\label{eq:S_matrix}
	    \end{equation}

	    \begin{widetext}
	     An explicit computation of the four matrix elements in the horizontal/vertical polarization ($H/V$) basis yields
	    	\begin{subequations}
	    		\begin{align}
	    			\mathcal{S}(\omega,\bar{\boldsymbol{\alpha}})_{HH(VV)}=&-1+\frac{i\kappa_{c}}{f(|\bar{\boldsymbol{\alpha}}|^2)}\left[m_\mathrm{eff}\left(\omega^{2}-\Omega_{m}^{2}+i\Gamma_{m}(\omega)\right)\left(\omega+\tilde{\Delta}+i\frac{\kappa}{2}\right)- G^{2}|\bar{a}_{V(H)}|^{2}\right],\\
	    			\mathcal{S}(\omega,\bar{\boldsymbol{\alpha}})_{VH(HV)}=&\frac{i}{f(|\bar{\boldsymbol{\alpha}}|^2)}\kappa_{c} G^{2}\bar{a}_{H(V)}^{*}\bar{a}_{V(H)},\label{eq:offdiag}
	    		\end{align}
	    	\end{subequations}
    	where $f(|\bar{\boldsymbol{\alpha}}|^2)=\left(\omega+\tilde{\Delta}+i\frac{\kappa}{2}\right)\left[m_\mathrm{eff}\left(\omega^{2}-\Omega_{m}^{2}+i\Gamma_{m}\omega\right)\left(\omega+\tilde{\Delta}+i\frac{\kappa}{2}\right)- G^{2}|\bar{\boldsymbol{\alpha}}|^2\right]$.
	    \end{widetext}
		The expressions  above capture arbitrary detuning conditions. In the main text, we consider a red-detuned cavity in the resolved-sideband limit ($\kappa\ll\Omega_m$), where the rotating-wave approximation ($\tilde{\Delta}\simeq-\Omega_m$),
		simplifies \autoref{eq:linear_H} to the excitation-conserving interaction $\hat{H}_\mathrm{eff}\simeq-\tilde{\Delta}\delta\hat{\textbf{a}}^{\dagger}\delta\hat{\textbf{a}}+\Omega_m\hat{b}^{\dagger}\hat{b}-g_0(\bar{\boldsymbol{\alpha}}\cdot\delta\hat{\textbf{a}}^{\dagger}\hat{b}+\mathrm{H.c.})$, using $\hat{x}=x_\mathrm{ZPF}(\hat{b}+\hat{b}^{\dagger})$ and defining $g_0=G/x_\mathrm{ZPF}$. In this limit, 
		\begin{align}
			\dot{\delta\hat{\textbf{a}}}=&-i(\Delta+i\kappa/2)\delta\mathbf{\hat{a}}+ig_{0}\bar{\boldsymbol{\alpha}}\hat{b} + \textbf{s}_\mathrm{in},
			\\\dot{\hat{b}}=&-i(\Omega_{m}+i\Gamma_{m}/2)\hat{b}+ig_{0}\bar{\boldsymbol{\alpha}}^{\dagger}\cdot\delta\hat{\textbf{a}}.
		\end{align}
		
		Hence, the $\mathcal{S}$-matrix expressions become equivalent to \autoref{eq:S_matrix} after the replacement $G\rightarrow g_0$ and
		\begin{equation}
			\chi_m(\omega)\rightarrow \chi_m^{\mathrm{RWA}}(\omega)=(-i(\Omega_{m}+\omega)+\Gamma_{m}/2)^{-1}.
		\end{equation}
		
		Note also that \autoref{eq:offdiag} clearly shows that the $\mathcal{S}$-matrix becomes nonreciprocal ($\mathcal{S}_{VH} \neq \mathcal{S}_{HV}$) when the steady-state cavity fields have a relative phase difference along two orthogonal axes, such as with the circularly-polarized pump used in the main text.
	
	\begin{figure*}[t]
			\centering
			\includegraphics[width=\linewidth]{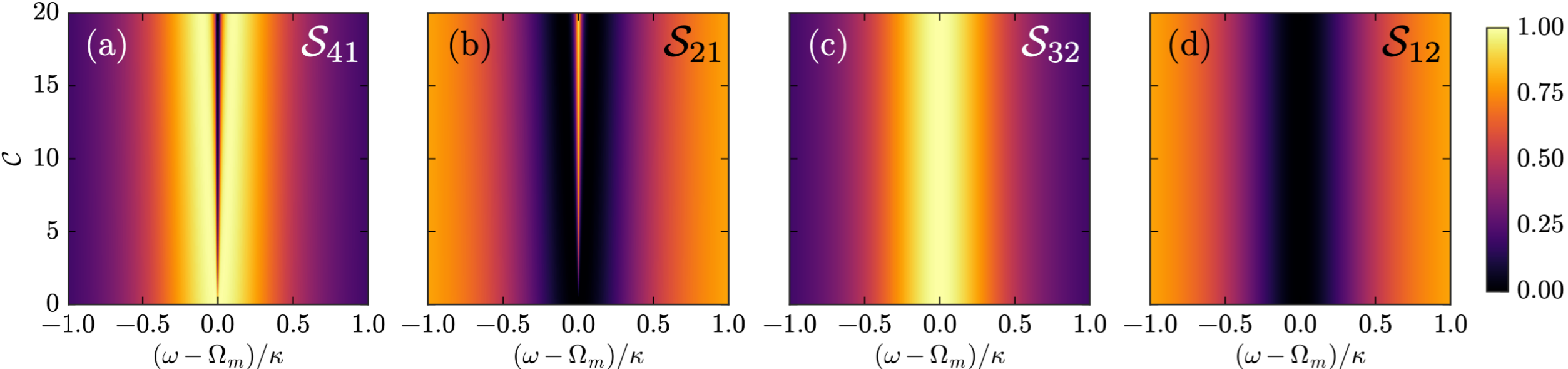}
			\caption{$\mathcal{S}$-matrix parameters for the circulator geometry sketeched in the main text with an RH control field. The H input probe fields become RH after passing through the QWP, and exhibit reflection on the mechanical sideband as the cooperativity increases, leaving through the V port on the same side. The V fields however pass through the resonator on resonance, independent of pump power. The coefficients $\mathcal{S}_{23}$, $\mathcal{S}_{43}$, $\mathcal{S}_{14}$, and $\mathcal{S}_{34}$ are equal to $\mathcal{S}_{41}$, $\mathcal{S}_{21}$, $\mathcal{S}_{32}$, and $\mathcal{S}_{12}$, respectively. For these plots, $\kappa=\kappa_{c}=2\pi\times5$ GHz. }
			\label{fig4}
	\end{figure*}
	
	\section{Geometrical action of the Scattering Matrix }
	
	In this section we offer a geometrical interpretation of the scattering matrix in \autoref{eq:S_matrix} within the high cooperativity limit ($\mathcal{C}=4g_{0}^{2}|\bar{\boldsymbol{\alpha}}|^{2}/(\Gamma_{m}\kappa)\gg1$) and  resonance ($\omega=\Omega_m$). The action of $\mathcal{S}$ on $\textbf{s}_\mathrm{in}$ is thus given by
	\begin{equation}
	\mathcal{S}(\Omega_{m},\bar{\boldsymbol{\alpha}})\textbf{s}_\mathrm{in}\overset{\mathcal{C}\gg1}{\simeq}-\textbf{s}_\mathrm{in}+\frac{2\kappa_{c}}{\kappa}\left(\textbf{s}_\mathrm{in}-\langle\textbf{e}_{\bar{\boldsymbol{\alpha}}},\textbf{s}_\mathrm{in}\rangle\textbf{e}_{\bar{\boldsymbol{\alpha}}}\right),\label{eq:S_matrix_geom}
	\end{equation}
	where $\textbf{e}_{\bar{\boldsymbol{\alpha}}}=\bar{\boldsymbol{\alpha}}/|\bar{\boldsymbol{\alpha}}|$ is a complex polarization vector for the control field. Neglecting the intrinsic loss channel, $\kappa\simeq\kappa_c$, and \autoref{eq:S_matrix_geom} shows the mapping implemented by $\mathcal{S}$ reflects the input vector over the (complex) control field. In order to determine how this mapping translates into a geometrical operation in the Bloch sphere, the control polarization vector is written as $\textbf{e}_{\bar{\boldsymbol{\alpha}}}=(\cos\theta_{\mathbf{k}}/2,e^{i\varphi_{\mathbf{k}}}\sin\theta_{\mathbf{k}}/2)$ and the $\mathcal{S}$-matrix  is expanded in the Pauli basis as
	\begin{equation}
		\mathcal{S}(\Omega_{m},\bar{\boldsymbol{\alpha}})=\mathbf{1}-2\textbf{e}_{\bar{\boldsymbol{\alpha}}}^{\dagger}\textbf{e}_{\bar{\boldsymbol{\alpha}}}=-\textbf{v}_{\bar{\boldsymbol{\alpha}}}\cdot\boldsymbol{\sigma},
	\end{equation}
	where  $\textbf{v}_{\bar{\boldsymbol{\alpha}}}=\left(\begin{array}{ccc}
	\sin\theta_{\mathbf{k}}\cos\varphi_{\bar{\boldsymbol{\alpha}}} & \sin\theta_{\mathbf{k}}\sin\varphi_{\bar{\boldsymbol{\alpha}}} & \cos\theta_{\bar{\boldsymbol{\alpha}}}\end{array}\right)$ is the 3-vector representing the control field in the Poincaré sphere and $\boldsymbol{\sigma}=(\sigma_x,\sigma_y,\sigma_z)$. This procedure effectively allows to switch from a (complex) $SU(2)$ representation to a (real) $O(3)$ representation, in which the scattering transfomation of the input state occurs via a similarity transform $	\mathcal{S}(\textbf{v}_\mathrm{in}\cdot\boldsymbol{\sigma})\mathcal{S}^{\dagger}$, which can be  expressed as 
	\begin{equation}
		\textbf{v}_\mathrm{out}\cdot\boldsymbol{\sigma}=\left[-\textbf{v}_\mathrm{in}+2\left(\textbf{v}_{\mathrm{in}}\cdot\mathbf{v}_{\theta,\varphi}\right)\textbf{v}_{\bar{\boldsymbol{\alpha}}}\right]\cdot\boldsymbol{\sigma}.
	\end{equation}
	Here we exploited the relation $ \left(\mathbf{a}\cdot\boldsymbol{\sigma}\right)\left(\mathbf{b}\cdot\boldsymbol{\sigma}\right)\left(\mathbf{a}\cdot\boldsymbol{\sigma}\right)=-\left(\mathbf{a}\cdot\boldsymbol{\sigma}\right)+2\left(\mathbf{a}\cdot\mathbf{b}\right)\left(\mathbf{b}\cdot\boldsymbol{\sigma}\right)$, for $\textbf{a},\textbf{b}\in\mathbb{R}^3$. Since we are free to choose the representation for $\boldsymbol{\sigma}$, the general expression 
	\begin{equation}
		\textbf{v}_\mathrm{out}=-\textbf{v}_\mathrm{in}+2\left(\textbf{v}_{\bar{\boldsymbol{\alpha}}}\cdot\textbf{v}_\mathrm{in}\right)\textbf{v}_{\bar{\boldsymbol{\alpha}}},\label{eq:ref}
	\end{equation}
	must hold. \autoref{eq:ref} is a particular case of the Rodrigues rotation formula for a real 3-vector around an axis $\textbf{v}_{\bar{\boldsymbol{\alpha}}}$ and angle $\epsilon=\pi$ \cite{Brockett2005}, given by $\mathbf{a}_{\mathrm{rot}}=\mathbf{a}\cos\epsilon+(\textbf{v}_{\bar{\boldsymbol{\alpha}}}\times\mathbf{a})\sin\epsilon+\textbf{v}_{\bar{\boldsymbol{\alpha}}}(\textbf{v}_{\bar{\boldsymbol{\alpha}}}\cdot\mathbf{a})(1-\cos\epsilon)=-\mathbf{a}+2(\textbf{v}_{\bar{\boldsymbol{\alpha}}}\cdot\mathbf{a})\textbf{v}_{\bar{\boldsymbol{\alpha}}}$. This formula outputs the rotated vector by decomposing the input into its components parallel and perpendicular to $\textbf{v}_{\bar{\boldsymbol{\alpha}}}$ and rotating only the perpendicular component.
	
	\hspace{0mm}
	\section{Circulator discussion}
	
	Utilizing polarization-dependent interactions, we can create a circulator with a CP pump and a two-sided cavity, as shown in the main text. The starting point consists of an unpumped, lossless cavity that transmits all fields on resonance and is thus reciprocal. An RHCP pump then induces reflection of for RHCP probe signals, with strength and bandwidth proportional to the cooperativity. We offer here a quick reminder that we define RHCP as a field component  $(\textbf{e}_H+i\textbf{e}_V)/\sqrt{2}$ without regard for propagation direction. With the setup of Fig. 3d, in the high cooperativity regime, the H input from both sides of the resonator is converted to RHCP by the QWP, reflected from the resonator, and converted to V by a second pass through the QWP on the same side. The V input instead can pass through the resonator as LHCP light, and passes through the opposite H port. Thus, on resonance, there is circulation between ports $1\rightarrow 2\rightarrow 3 \rightarrow 4\rightarrow 1$, shown clearly in  \autoref{fig4}.

\bibliographystyle{apsrev4-1}
\bibliography{OM_BFE_SM}